%% file: MAIN_galaxy.tex
\documentclass[fleqn,usenatbib]{rasti}

\usepackage{newtxtext,newtxmath}
\usepackage[T1]{fontenc}

\DeclareRobustCommand{\VAN}[3]{#2}
\let\VANthebibliography\thebibliography
\def\thebibliography{\DeclareRobustCommand{\VAN}[3]{##3}\VANthebibliography}

\usepackage{graphicx}
\usepackage{amsmath}	

\usepackage{caption} 

\usepackage{color,soul}
\usepackage{natbib} 
\makeatletter
\xpatchcmd\NAT@citex
 {%
  \@citea\NAT@hyper@{%
    \NAT@nmfmt{\NAT@nm}%
    \hyper@natlinkbreak{\NAT@aysep\NAT@spacechar}{\@citeb\@extra@b@citeb}%
    \NAT@date
  }%
 }
 {%
  \@citea
  \NAT@nmfmt{\NAT@nm}%
  \NAT@aysep\NAT@spacechar
  \NAT@hyper@{\NAT@date}%
 }
 {}{}
\xpatchcmd\NAT@citex
 {%
  \@citea\NAT@hyper@{%
    \NAT@nmfmt{\NAT@nm}%
    \hyper@natlinkbreak{\NAT@spacechar\NAT@@open\if*#1*\else#1\NAT@spacechar\fi}%
    {\@citeb\@extra@b@citeb}%
    \NAT@date
  }%
 }
 {
  \@citea
    \NAT@nmfmt{\NAT@nm}%
    \NAT@spacechar\NAT@@open\if*#1*\else#1\NAT@spacechar\fi
    \NAT@hyper@{\NAT@date}%
 }
 {}{}
\makeatother

\usepackage{url}

\usepackage{xcolor}
\usepackage{xpatch}

\usepackage{hyperref}
\hypersetup{
    colorlinks=true,
    citecolor=cyan,
    urlcolor=cyan,
    linkcolor=black
    }

\usepackage{orcidlink} 

\makeatletter
\renewcommand{\footnoterule}{%
  \kern -3pt
  \hrule width 2in
  \kern 2.6pt
}

\makeatother

\title[Galaxy Orientation Angles]{Inclinations and Position Angles for Disc Galaxies in the SGA sample}

\author[M. H. Martinez et al.]{
Megan H. Martinez$^{\orcidlink{0009-0003-5203-8051}{1,2}}$\thanks{E-mail: meganhm@tcd.ie},
Michael S. Petersen$^{\orcidlink{0000-0003-1517-3935}{2}}$,
Carrie Filion$^{\orcidlink{0000-0001-5522-5029}3}$,
Rashid Yaaqib$^{\orcidlink{0009-0003-9063-1382}2,4}$
and Claire Larson$^{\orcidlink{0009-0006-7814-7901}2}$
\\
$^{1}$Trinity College Dublin, The University of Dublin, College Green, Dublin 2, D02 PN40, Ireland \\
$^{2}$Institute for Astronomy, University of Edinburgh, Royal Observatory, Blackford Hill, Edinburgh EH9 3HJ, UK\\
$^{3}$Center for Computational Astrophysics, Flatiron Institute, New York, NY 10010\\
$^{4}$Department of Physics, United Arab Emirates University, Al Ain, Abu Dhabi, UAE\\
}

\date{Accepted XXX. Received YYY; in original form ZZZ}

\pubyear{\the\year{}}

\begin{document}
\label{firstpage}
\pagerange{\pageref{firstpage}--\pageref{lastpage}}
\maketitle
\input{0Numbers}
\begin{abstract}
We present a data-driven method for determining the inclination and position angle (PA) of disc galaxies using a Fourier-Laguerre basis decomposition of imaging data. We define a dimensionless metric, $\eta$, that characterises the ratio of the quadrupole and monopole coefficients in the Fourier-Laguerre basis function expansion. This metric serves as a robust measure which is related to the inclination of a galaxy. We find an empirical relationship between $\eta$ and inclination which is agnostic to the galaxy morphology. The PA is derived directly from the phase of the quadrupolar Fourier-Laguerre functions.  Across a benchmark sample of galaxies, the method reproduces published inclination and PA values to within a median of $10^\circ$ and $5^\circ$, respectively, while also demonstrating essentially zero catastrophic failures. Applying this pipeline to galaxies from the Siena Galaxy Atlas (SGA), we report measurements of $\eta$, scale length and PA for three different bands of \sgaused\ disc galaxies. 
Our computationally inexpensive technique automates parametrisation analysis and returns reproducible results for large surveys. We release a Python package ready for application to next generation surveys.
\end{abstract}




\section{Introduction}
\input{2Intro}

\section{Methodology}
\input{3Methodology}

\section{Data expansions}
\input{4Data}

\section{Results}
\label{sec:results}
\input{5Results}

\section{Conclusions}
\label{sec:concl}
We provide inclination, PA and scale length measurements for \sgaused\ SGA disc galaxies derived entirely from imaging data using our Fourier–Laguerre expansion method. The dimensionless parameter $\eta$, defined as the ratio between different orders of the $m=2$ sine and cosine terms in the Fourier-Laguerre basis pairs, correlates strongly with previously measured galaxy axial ratios (\textit{b/a}), but offer a more physically motivated measure of galaxy inclination. These results reproduce literature values with a typical scatter of approximately 10$^\circ$ in inclination and $5^\circ$ in PA. Moreover, this method provides reliable and reproducible estimates of galaxy orientation, independent of morphology or apparent size.

Future extensions of this work will focus on interpreting Fourier-Laguerre coefficients for quantitative morphology determination. Understanding the effect of inclination on the coefficients is a critical first step. The higher order coefficients can naturally capture fine structure details such as bars, spirals and arms owing to the basis capturing both radial and azimuthal structure. Applications are therefore not limited to inclinations and PAs, but also include a data-driven approach to study the geometry and evolution of disc galaxies and classifying their entire morphologies. Future comparisons to simulations \citep[e.g.][]{Reddish2022,Petersen2024,Tep2025} will also be enabled by correcting for inclination effects.

The method presented here is reproducible, computationally inexpensive, morphology-independent, and comfortably extended to future LSST- or other survey-scale datasets. It is ideally suited where automated, morphology-agnostic processing is essential.

\section*{Acknowledgements}
MHM acknowledges the support of the University of Edinburgh Vacation Research Programme.
MSP acknowledges support from a UKRI Stephen Hawking Fellowship. The authors of this paper utilized the following software: \textsc{numpy} \citep{harris2020array}, \textsc{scipy} \citep{2020SciPy-NMeth}, \textsc{astropy} \citep{2018AJ....156..123A}, \textsc{hdf5} \citep{fortner1998hdf} and \textsc{matplotlib} \citep{Hunter:2007}. We acknowledge the usage of the HyperLEDA database (\url{http://leda.univ-lyon1.fr}).\\

\section*{Data Availability}

All SGA data is available at \url{https://portal.nersc.gov/project/cosmo/temp/ioannis/sga-dr9/}, with summary data available in \url{https://portal.nersc.gov/project/cosmo/data/sga/2020/SGA-2020.fits}. The analysis tools are available as part of the {\tt ObservationalExpansions} GitHub organisation.\footnote{\url{https://github.com/ObservationalExpansions}}



\bibliographystyle{rasti}
\bibliography{refs} 




\appendix
\section{Mathematical Background}\label{Appendix_Maths}

Observational evidence suggests that the surface brightness profile of disc galaxies decreases exponentially \citep{Freeman:1970mx}. We can then write the surface density, averaging over the azimuthal angle $(\phi)$, as 
\begin{equation}
    \Sigma(R,\phi)=\Sigma(R)=\Sigma_0\exp{\left(-\frac{R}{a}\right)},
    \label{surfacedensity}
\end{equation}
where $\Sigma_0$ is the central surface density, $a$ is the \textit{scale length} of the disc and, as the system is in polar coordinates in a projected two-dimensional plane, the radius of the position of the point ($R$) is given by $R=\sqrt{X^2+Y^2}$ and the angular position ($\phi$) by $\phi=\arctan(Y/X)$.
Since imaging and pixel data is being used, $\Sigma_0$ can be taken as the maximum central pixel. Here, $X$ and $Y$ are the pixel coordinates in the image, rather than point position, however both can be considered.
\par Like other methods with alternative special functions serving as a basis to reconstruct the galaxy \citep{rene_diffbasis}, the basis, in this case a basis formed with Fourier series and Laguerre polynomials, reflects the intensity and position of the points/pixels within the image and reconstructs the image as an approximation of these qualities.

\subsection{Laguerre Polynomials}
The radial and azimuthal components correspond to Laguerre polynomials and Fourier series, respectively \citep{PetersenWeinberg}.
For index $n$, the Laguerre basis, in terms of associated Laguerre polynomials, $L_n^\alpha$, is given by,
\begin{equation}
\label{laguerre}
    \mathcal{L}_n(R)=\frac{2}{a\sqrt{n+1}}\exp{\left(-\frac{R}{a}\right)}L_n^1\left(\frac{2R}{a}\right),
\end{equation}
where $L_n^1$ is the associated Laguerre polynomial for $\alpha=1$. In this case, $R$ is the radius and $a$ is the scale length of the galaxy.
\subsection{Combining with Fourier Series}
Introducing the azimuthal dependence, with index $m$, via a Fourier series, one can combine the series with the Laguerre basis from Equation~(\ref{laguerre}) to form the Fourier-Laguerre pairs as a set of two-dimensional basis functions given by $\mathcal{L}_n(R)\cos{m\phi}$ and $\mathcal{L}_n(R)\sin{m\phi}$ \citep{ReichardFourier,PetersenWeinberg, Ananya}. Projecting the surface density from Equation~(\ref{surfacedensity}) into the basis functions, a pair of coefficients corresponding to sine and cosine terms are returned as a discrete sum given by,
\begin{equation}
    \begin{split}
        \hat{c}_{mn}=\sum_x\sum_y\lambda_m(R_{xy},\phi_{xy})\mathcal{L}_n(R_{xy})\cos(m\phi_{xy}),\\
        \hat{s}_{mn}=\sum_x\sum_y\lambda_m(R_{xy},\phi_{xy})\mathcal{L}_n(R_{xy})\sin(m\phi_{xy}).
    \end{split}  
\end{equation}
Here, $\lambda_{m}$ is the angular normalisation given by,
\begin{equation*}
    \lambda_{m}=\begin{cases} 
        \frac{1}{\pi} & \text{if } m = 0, \\
        \frac{2}{\pi} & \text{if } m > 0.
    \end{cases}
\end{equation*}
A reconstruction of the model surface density profile is made by summing over the coefficients in the following way,
\begin{equation}\label{eqn:recon}
    \hat{\Sigma}_\text{disc}(R,\phi)=\sum_m\sum_n\left(\hat{c}_{mn}\cos(m\phi)+\hat{s}_{mn}\sin(m\phi)\right)\mathcal{L}_n(R).
\end{equation}
In order to convert the sine and cosine coefficients into a metric corresponding to inclination and PA values, one can consider a matrix formed from the amplitude of the coefficients. This can be represented as,
\begin{equation}\label{matA}
\bf{A}=\begin{bmatrix}
    A_{00} & A_{10} & A_{20} &\dots&A_{m_{\text{max}}0}\\
    A_{01}& A_{11} &&\dots\\
    A_{02}&& \ddots\\
    \vdots&\vdots\\
    A_{0n_{\text{max}}}&&&\dots&
    A_{m_{\text{max}}n_{\text{max}}}\\
    \end{bmatrix},
\end{equation}
where each $A_{mn}=\sqrt{\hat{c}_{mn}^2+\hat{s}_{mn}^2}$. The matrix $\bf{A}$ has rows of increasing $n$ (Laguerre) order and columns of increasing $m$ (Fourier) order. The components of $\bf{A}$ are the \textit{coefficients} for each order of the corresponding Fourier-Laguerre pairs.

\section{Calibration and Recommended Inversion Relation}\label{Appendix_recommendedIncl}

In this Appendix, we use idealised exponential disc models to test the Fourier-Laguerre methodology and validate relationships between inclination and position angle with coefficient indicators. We realise exponential discs by drawing points from an exponential distribution and randomly selecting a position angle for each point. We then `observe' the galaxy by making a histogram of the point locations. 

The software is implemented in {\tt discmodel}, available on GitHub.\footnote{\url{https://github.com/ObservationalExpansions/discmodel}}

\subsection{Point vs. pixel expansions}

Ideally, one would apply a Fourier-Laguerre expansion to a galaxy in terms of each point and mass in the galaxy and measure the positions to infinitesimal point particles \citep{2005Natur.435..629S}, however when working with data and images, they are in pixel format. The only way to check if this method remains consistent for the so-called pixel expansions or point expansions is to expand the disc model galaxy both for points and for an average of points (pixels) and verify a consistent result. Running both approaches for a disc model galaxy for an array of inclination and PA values yields consistent trends in the recovered $\eta$ relation. The dependence of the expansion parameters on galaxy orientation remains robust whether one works with pixel expansions or point expansions. This ensures that the pixel-based expansion formula can be reliably applied to real imaging data, where only pixel intensities are available. Thus, the verification on model disc galaxies demonstrates that the pixel expansion is not merely a numerical convenience but a theoretically solid approximation of the underlying point-particle expansion.

\subsection{Empirical relation and calibration}

\begin{figure}
    \centering
    \includegraphics[width=1\linewidth]{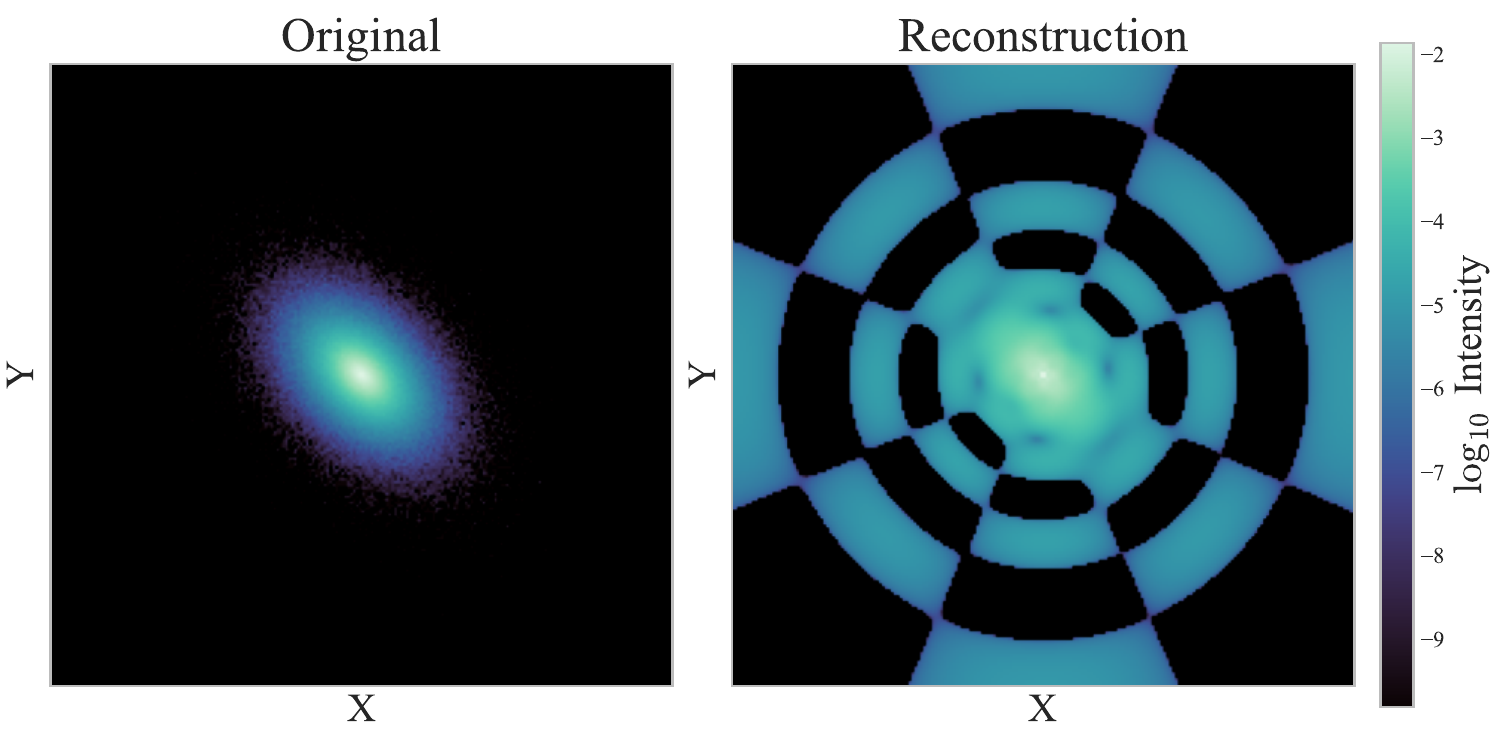}

    \caption{(i) Original (left) and (ii) Laguerre reconstruction (right) of a simulated disc galaxy inclined at $50^\circ$ with PA $45^\circ$.}
    \label{fig:galaxy50_45}
\end{figure}

Performing a Fourier-Laguerre expansion on a galaxy returns a coefficient matrix with $m_{\rm max}\times n_{\rm max}$ values. Following the mathematical details in Appendix~\ref{Appendix_Maths}, the coefficients can be used to create a reconstruction of a galaxy (see Equation~\ref{eqn:recon}). Taking a disc galaxy inclined at $i=50^\circ$ and rotated at ${\rm PA}=45^\circ$ as an example, one can expand and compare the reconstruction with the original disc in Figure~\ref{fig:galaxy50_45}. One can draw clear parallels between the original disc and the shape at the centre of the reconstruction. This particular reconstruction was made with $n_\text{max}=10$ and $m_\text{max}=6$. At a first glance, one can observe the change in intensity of the pixels as they trace out the elliptical shape of the disc. The block structure seen as arc sections in the increasing radial direction display the structure of the Fourier-Laguerre pairs working together to reconstruct the image. Where there are no pixels to represent in the expansion, their natural form with equal intensity is visible. For the inclination and PA measurements, only the pixels representing the galaxy and their intensities are required.

\par With the intention of finding an invertible relation to output an inclination, the $\eta$ value for incrementally inclining galaxy models are stored along with their corresponding inclination. This paper presents suggested conversion approximations to perform on the reported $\eta$ values. 
As a general approximation, a cosine curve can approximate inclinations. To achieve a closer match to the perfect exponential disc, a relation involving an exponential can achieve more accurate values for higher inclinations. The formula for $\eta$ is given by,
\begin{equation}\label{eta_function}
    \eta=A\cos(Bi+C)+D+E\exp(Fi),
\end{equation}
where $i$ is the inclination and $A,B,C,D$ and $F$ are fit parameters. This function would require numerics and a root solver to be efficiently applied as an inversion. 
While values for $\eta$ are reported and some matching literature values are retrieved, for galaxies which in general do not resemble an exponential disc, a more complicated relation may need to be derived. 

We define the inclination as a function of $\eta$, based on the postulate that a bijective mapping between the inclination $i$ and $\eta$ exists. This is followed by the specifics to selecting one particular conversion or another. 

\par For the purposes of this paper, we detail an example which implements the following conversion relation. Equation~(\ref{eta_function}) was inverted with numerical methods and a root solver to best identify the corresponding inclination given an $\eta$ value. The best parameters found for images larger than 256$\times$256 pixels were A=-0.243, B=-2.275, C=8.593, D=-1.000, E=0.938 and F=0.010 to three decimal places.

An alterative approach to a sturdy inversion relation involves fitting a curve proportional to $\cos^2(i)$ which intuitively matches our understanding of the expansion pairs and their physical representation. Fitting a curve with the form,
\begin{equation}
    \eta = A\cos^2(Bi+C)+D,
\end{equation}
with $A,B,C$ and $D$ being fit parameters while also ensuring $\eta(0)=\eta_\text{min}$ and $\eta(90^\circ)=\eta_\text{max}$ allows for a bijection and inversion to find inclination to take place. Here, $\eta_\text{min/max}$ are the minimum and maximum simulated eta values for the generated inclined discs. Forcing boundary conditions enforces a reasonable gauge for intermediate values while following the same curvature of the data points. This means that $B$ becomes only free variable, which we determine to be approximately 0.2 with a Python fitting program.

\section{Pipeline}\label{Appendix_Pipeline}

\begin{figure*}
    \centering
    \includegraphics[width=1\textwidth]{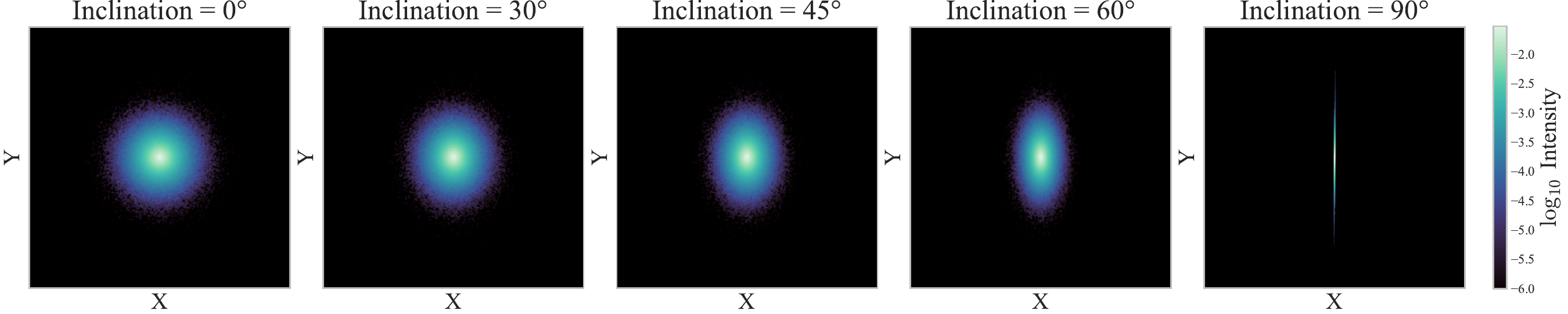}
    \caption{Exponential disc model inclining at angles $0^\circ,30^\circ,45^\circ,60^\circ,90^\circ$.}
    \label{fig:GalaxyIncl}
\end{figure*}

\begin{figure*}
    
    \centering
    \includegraphics[width=1\textwidth]{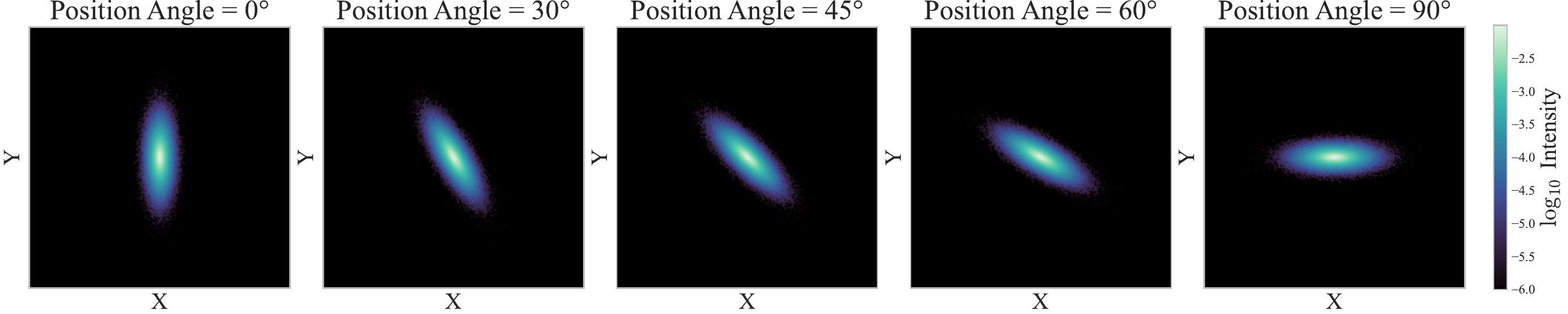}
    \caption{Exponential disc model inclined at $70^\circ$ with PAs of $0^\circ,30^\circ,45^\circ,60^\circ,90^\circ$.}
    \label{fig:GalaxyPA}
\end{figure*}

Motivated by the \texttt{flex} pipeline created for the Fourier-Laguerre expansions \citep{Ananya}, the expansions following the equations described above could be applied to any galaxy image. In this Appendix, we provide demonstrations of how {\tt discmodel} and {\tt flex} can be used in tandem to exlore idealised models and measurements. First, a perfect exponential disc was generated with the new \texttt{discmodel} pipeline, whose properties such as inclination, PA, scale length and mass density could be modified. A model was made of a disc galaxy that was gradually inclined from $0^\circ-90^\circ$ and rotated through $0^\circ-180^\circ$.
Using {\tt discmodel} to generate images of disc galaxies at various inclinations and PAs produces Figure~\ref{fig:GalaxyIncl}~ and Figure~\ref{fig:GalaxyPA}, maintaining the conventions for the measurements. This demonstrates how the basis should account for a disc transitioning to a line as a perfect disc in two-dimensions inclines to a razor thin line. 

\bsp	
\label{lastpage}
\end{document}

%% file: 0Numbers.tex
\newcommand{\sgatotal}{383,620}
\newcommand{\sgaselected}{152,109}
\newcommand{\sgaused}{133,942}
\newcommand{\sgaflag}{120,650}

%% file: 2Intro.tex
Understanding galaxies and their morphologies has been at the forefront of astrophysics since the ancient philosophers. Following the first image of Andromeda in \citet{Roberts1888}, of what was then known as `The Great Nebula of Andromeda', telescopes and astrophotography have advanced to capture images of millions of distinct galaxies \citep{grant2024andromeda}. Telescopes, such as the \textit{Hubble Space Telescope} and the \textit{James Webb Space Telescope (JWST)}, along with all-sky surveys such as the \textit{Sloan Digital Sky Survey (SDSS)} and the \textit{Legacy Survey of Space and Time (LSST)}, have contributed to finding and classifying the morphologies of galaxies. The Hubble sequence remains the standard framework for classifying spiral and barred galaxies \citep{Hubble}. 

Another feature of galaxies, which can be studied as a direct observable, is their orientation. The two main orientation angles are position angle (PA) and the inclination angle. The PA is the fixed rotation of the galaxy in the plane of the sky. Inclination is measured as the angle through which a galaxy tilts into, along the line of sight, normal to the plane of the sky. 
Disc galaxies' evolution and formation have been studied in great detail, with a major result being the fact that their stellar surface density is characterised by an exponential radial profile \citep{Freeman:1970mx}. Combining the exponential surface intensity profile with PA and inclination defines a standard model of a disc galaxy. Thus, PA and inclination are highly meaningful measurements for disc galaxies, and are important for determining galaxy properties.

The motivation to study inclination and PA values of disc galaxies first started in the early 20$\text{th}$ century, when Hubble considered inclined disc galaxies projected as ellipses and found the ratio between the minor and major axes, or the \textit{b/a} ratio \citep{Hubble}. He recognised that this apparent flattening was primarily a geometric effect: as a circular disc is tilted with respect to the line of sight, its projection becomes more elliptical, causing the b/a ratio to decrease systematically with increasing inclination. This provided an observational link between the measured axis ratio and a galaxy’s intrinsic in-plane inclination. Some went further to analyse the isophotal twisting and other properties which could lead to inconsistencies with this ratio \citep{benacchioincl,binny}. In addition to analysing galaxies and their ellipticity from distributions of data \citep{bovera}, it is convenient to implement approximations and compression of images. Extensive effort has been devoted to finding a way to automate the classification of morphologies of galaxies \citep{classificationsJangren,ANewApproach,ferrari2015morfometrykanewway}, including techniques of approximating and compressing data and images \citep{rene_diffbasis,Kelly_2004,fraix2015,Semenov_2025} as well as simulations \citep{Petersen2016, Petersen2019,Petersen2021,Johnson2023,Arora2025,Hunt2025,DarraghFord2025}.
Following the methods pioneered previously \citep{PetersenWeinberg,Ananya}, the use of  Fourier-Laguerre basis can reproducibly reduce the quantity of data being analysed and construct an approximate galaxy image. This method takes advantage of Fourier series and Laguerre polynomials as a set of orthogonal basis functions. Represented as weights corresponding to each order of the series/polynomial, the galaxy can be decomposed into a handful of terms. The image can then be approximated by re-projecting these weights into the basis functions. However, the weights themselves are meaningful and can be analysed directly. 
What follows in this work is a developed and evolved technique which inherits the Fourier-Laguerre basis expansion method (referred to as {\tt flex}). 

Incentivised by modern techniques which complement the traditional axis ratio measurement \textit{b/a}, we focus on metrics derived from the Fourier-Laguerre expansion to provide a systemic parametrisation of the galaxies \citep{Cappellari_2008,Maller_2009}. Incorrect orientation estimates can bias derived quantities such as rotational velocity, angular momentum, and stellar mass distributions \citep{Courteau_1997,hall}. Therefore, developing a robust and reproducible technique is crucial for studying galactic dynamics and classifying large surveys of data. Understanding the orientation and morphology of disc galaxies is essential for studying galaxy kinematics, formation, history, and environmental interactions. 

The utility of combining Fourier series and Laguerre polynomials is motivated by work that has been done previously, including work demonstrating how Fourier series azimuthal orders \citep{Zaritsky_1997,ReichardFourier} can be used to measure lopsidedness. Combining Fourier azimuthal descriptions together with a similar orthogonal description of radial order structure (Laguerre polynomials), one can approximate a galaxy in terms of the orders of the basis functions \citep{PetersenWeinberg,Ananya}. We present a new method, powered by \texttt{flex}, for measuring inclination values and PA values for a catalogue of galaxies. The main products are a dimensionless metric, $\eta$ -- which can be converted into inclination with calibrated inversion relations, a PA value, and the scale length of the galaxy.
This technique requires only a galaxy image and compresses the data into a compact set of coefficients. This technique aims to resolve inconsistencies in published angle measurements and provide parametrisations for galaxies in future datasets. Our publicly accessible code built with Python can be found in \url{https://github.com/meganhmartinez/Galaxy-Morphology}.
A considerable difference of the Fourier-Laguerre expansion recovered orientation relative to model-based metrics is that it is agnostic to the fact that the image is a galaxy. {\tt flex} does not fit explicit parametric forms for asymmetries, spirals, bars, lopsidedness or gas distribution, but these properties can be derived from the weights. This advantage allows for the same pipeline to be applied to a collection of galaxies and return appropriate inclination and PA values. 

In this work, we measure the inclinations, PAs, and scale lengths for a large sample of galaxies drawn from the \textit{Siena Galaxy Atlas} \citep[SGA;][]{moustakas2023sienagalaxyatlas2020}. SGA provides reprocessed imaging of nearby galaxies from the \textit{Dark Energy Spectroscopic Instrument} (DESI) Legacy Surveys, combining data from The Dark Energy Camera Legacy Survey (DECaLS) \footnote{\url{https://www.legacysurvey.org/decamls/}}, The Beijing-Arizona Sky Survey (BASS)\footnote{\url{https://www.legacysurvey.org/bass/}} and The Mayall z-band Legacy Survey (MzLS)\footnote{\url{https://www.legacysurvey.org/mzls/}}, to achieve consistent photometric calibration and improved background subtraction. Its goal was to produce accurate surface brightness profiles, shape measurements and photometric parameters for bright galaxies to support target selection and calibration for DESI. The SGA dataset was selected for our data due to its uniform, high-resolution imaging, as well as its morphology classification, allowing for an ideal large-scale test. 

Our Fourier-Laguerre technique offers a fully data-driven alternative to measuring orientation angles. Moreover, its stability and flexibility enables a simple injection into large datasets from (e.g.) SDSS, JWST or LSST without requiring prior morphological-based classification or training on specific galaxy types, complementing other quantitative morphology efforts \citep[e.g.][]{statmorph}. The paper is organised as follows. Section~\ref{sec:methodology} introduces the Fourier–Laguerre framework, Section~\ref{sec:dataexp} details the application to the SGA galaxies, Section~\ref{sec:results} compares our results with established catalogues, and Section~\ref{sec:concl} concludes with a summary and outlook.

%% file: 3Methodology.tex
\label{sec:methodology}
\subsection{Fourier-Laguerre expansions}

Following methodology introduced in \citet{PetersenWeinberg} and \citet{Ananya}, we apply a Fourier-Laguerre expansion to infer galaxy properties. The full details of the expansion are presented in Appendix~\ref{Appendix_Maths}. Briefly, we represent the light distribution of a galaxy as a sum of two-dimensional orthogonal basis functions in radial and azimuthal coordinates $(r,\phi)$, where the radial dependence is described by Laguerre polynomials, $\mathcal{L}_n(r)$, and the azimuthal dependence is described in Fourier harmonics, $\exp(im\phi)=\cos(m\phi)+i\sin(m\phi)$. The integers, $n$ and $m$, index the basis functions, where ($n,m$) are the radial and azimuthal orders respectively. By analysing the contributions to the light distribution from different combinations of functions, we can measure different properties of the galaxies, such as orientation.
For the purposes of this paper, we are interested in using the contributions from different functions to determine the inclination and PA. We refer to the weights specifying the contributions of different functions as {\it coefficients}.

The calculation of the Fourier-Laguerre coefficients is performed by {\tt flex},\footnote{\url{https://github.com/ObservationalExpansions/flex} {\tt v1.0.0}} a purpose-built basis function expansion driver for efficient two-dimensional Fourier-Laguerre basis function calculations. The Laguerre functions require a scale length as a parameter. In practice, we estimate this from images of galaxies by fitting an exponential function to the surface brightness profile. If the galaxy is a perfect exponential disc, the $m=0$ Laguerre term is exactly the surface brightness profile of the galaxy, highlighting the motivation for its use. The optimal scale length is found by minimising residuals between the model and an azimuthally-averaged observed surface-brightness profile for each galaxy, ensuring that the Laguerre radial term captures the exponential decline without over-fitting noise. In detail, we accept the SGA-reported (RA, Dec) as the centre of the galaxy and fit the azimuthally-averaged surface brightness profile between the centre and the radius at which the pixels are dominated by the background of the image (measured directly from a blank patch of the image).

\subsection{Coefficient ratios and their role in determining orientation angles}

To determine inclination, we are interested in coefficients that encode the galaxy's overall structure and shape. Considering only azimuthal dependence, a Fourier expansion of a perfect, face-on, featureless disc is described completely with only axisymmetric monopole terms ($m=0$). However, the Fourier basis forms a complete set which can describe the system from a disc to a straight line. For the purposes of galaxy viewing angles, as the disc inclines and appears as an ellipse in projections, additional even $m>0$ Fourier orders must be used to represent the tilted ellipse. The leading term corresponding to this representation of an inclined disc is the $m=2$ Fourier order.

Following the details in Appendix~\ref{Appendix_Maths}, we define $\eta$, a metric which will later be employed in finding the inclination, as the total contribution from the $m=2$ Fourier modes relative to the $m=0$ Fourier mode
\begin{equation}\label{eta}
    \eta=\frac{\sqrt{\sum_{n=0}^{n_\text{max}}
    \left(\hat{c}_{m=2,n}^2+\hat{s}_{m=2,n}^2\right)}}{\sqrt{\sum_{n=0}^{n_\text{max}}\hat{c}^2_{m=0,n}}}, 
\end{equation}
where $\hat{s}$ and $\hat{c}$ correspond to sine and cosine terms.\footnote{Discrete summations are denoted by the hat ($\hat{c} $ and $\hat{s}$), to distinguish from the purely theoretical continuous basis measurement, cf. \citet{PetersenEXP}.}
By summing over the columns from $n=0$ to some value $n=n_\text{max}$, where one has some flexibility with the choice in maximum $n$ order, the result is a dimensionless parameter which encodes this orientation. In order to limit the sum, after some testing we find that an $n_\text{max}$ value of 10 is an appropriate constraining balance between approximating the light distribution well and over-fitting. This choice provides a consistent and accurate representation of the surface density of the galaxy. Given this representation of the galaxy, $\eta$ becomes an inexpensive method of representing the information with which one can determine the inclination.

In Appendix~\ref{Appendix_recommendedIncl}, we derive an empirical relationship between $\eta$ and inclination using a perfectly thin and featureless disc galaxy. This relationship is used throughout the paper to convert measured $\eta$ values to inclinations. While this calibration provides a robust mapping for discs resembling an ideal exponential profile, the framework is flexible. Alternative functional forms can be substituted if applied to datasets with different morphologies, wavelength regimes or image resolutions and noise to convert between $\eta$ and conventional inclination measurements.

Similarly, we use the information in the phase angle of the Fourier expansion to determine the position angle of the galaxy,
\begin{equation}\label{vartheta}
    \vartheta=\arctan\left(\frac{\hat{s}_{m=2,n=0}}{\hat{c}_{m=2,n=0}}\right).
\end{equation}
As above, $\hat{s}$ and $\hat{c}$ correspond to the sine and cosine \textit{coefficients}.
By enforcing convention of measurement via the major axis PA (North Eastwards) increasing anti-clockwise, as defined in the \textit{HyperLEDA} catalogue, the PA is therefore retrieved by computing
\begin{equation}\label{PAeqn}
    {\mathrm{PA}}=\left(\frac{\vartheta}{m}\right)\left(\frac{180^\circ}{\pi}\right) + 90^\circ.
\end{equation}
The factor of $m$ appears owing to the multiplicity of the Fourier expansion, and for our calculations is always $m=2$. 
With Equation~(\ref{eta}) and Equation~(\ref{PAeqn}), we proceed to implement them into the expanded and reconstructed galaxies to find a relation for the inclination values and find the PA values.

The measurement of $\eta$ and PA are performed by \texttt{galaxymorphology}%
\footnote{
\url{https://github.com/ObservationalExpansions/Galaxy-Morphology}
\texttt{v1.1.0}},
a Python package that takes an image as the input and returns structural measurements.

%% file: 4Data.tex
\label{sec:dataexp}
In this section, we cross-checked the initial results derived from disc galaxies generated by the \texttt{discmodel} package \footnote{ \url{https://github.com/ObservationalExpansions/discmodel} \texttt{v1.0.0}}
against a `golden sample' of galaxies for which inclination has been re-derived using a combination of isophotal fitting and gas dynamics in \citet{GalaxyInclPaper} in Section~\ref{subsec:goldensample}. 
We summarise our sample of disc-identified galaxies drawn from the SGA catalogue and describe the {\tt galaxymorphology}-derived measurements in Section~\ref{subsec:allsga}. 
With the intention of measuring inclination, PA and to quality check the galaxies and manage uncertainties,
each selected galaxy is run through the {\tt galaxymorphology} function to measure the expansion scale lengths and obtain $\eta$ and PA values for a grid of the {\it grz} filters and a range of prefactors for the Laguerre scale lengths ($[0.75,1.0,1.5,2.0,2.5,3.0]$). As the inclination should agree independently of observed band, and should be insensitive to the expansion scale length within reasonable prefactor scale limits, we use the combination filter-and-scale length grid to define two quality flags. The first flag is the mean difference in $\eta$ between each combination of the three bands, and the second is the standard deviation in $\eta$ between the different scale lengths. By inspection of the grid of $\eta$ values for all galaxies in the sample, we determine empirical thresholds of $\sigma_{\eta,~{\rm band~difference}}=0.06$ and $\sigma_{\eta,~{\rm scale~difference}}=0.06$, but find in practice that small changes do not significantly affect the results. Lastly, we define a `contamination' threshold by inspecting the dipole distortion of each galaxy by computing the equivalent of $\eta$ for $m=1$,
\begin{equation}\label{zeta}
    \zeta=\frac{\sqrt{\sum_n
    \left(\hat{c}_{m=1,n}^2+\hat{s}_{m=1,n}^2\right)}}{\sqrt{\sum_n\hat{c}^2_{m=0,n}}},
\end{equation}
and defining a contamination threshold of ${\rm max}\left(\zeta_{g},\zeta_r,\zeta_z\right)=0.5$. This threshold captures expansions that are affected by either foreground stars or background galaxies. We combine these different quality flags into a single bit flag, where galaxies failing $\sigma_{\eta,~{\rm band~difference}}$ are given a `1', galaxies failing $\sigma_{\eta,~{\rm scale~difference}}$ are given a `2', and galaxies failing the $\zeta$ threshold are given a `4'. The flags are then summed into one indicator, {\tt quality\_flag}. We recommend using only galaxies with {\tt quality\_flag}$<=3$, though other flag values can be used as an indicator of physically interesting features, e.g. strong disturbances or companions resulting in deviations from an exponential profile.

\subsection{`Golden Sample'}
\label{subsec:goldensample}

Motivated by the goal of validating the performance of our Fourier–Laguerre expansion method, we define a \textit{golden sample} of galaxies to be a subset of thirty-nine galaxies sourced from \citet{GalaxyInclPaper} and cross referenced with \textit{HyperLEDA}. Their work was selected as it presents a catalogue for a subset of galaxies and aimed to improve the measurements provided by \textit{HyperLEDA}.
These objects provide a benchmark for testing the accuracy and reproducibility of the technique.

For each galaxy, we applied the same {\tt galaxymorphology} pipeline as used later for the larger SGA dataset, performing Fourier–Laguerre expansions in the \textit{g}, \textit{r}, and \textit{z} bands. 
The recovered inclination and PA estimates were consistent across filters, with minor inter-band variations in both inclination and PA. These measurements provide an independent test of the stability of the pipeline and highlight the empirical recommended $\eta$-inclination calibration presented in Appendix~\ref{Appendix_recommendedIncl}.

A detailed comparison between our results and publicly available catalogues, including \textit{HyperLEDA}, is presented in Section~\ref{subsec:gsres}, where we evaluate the accuracy and reproducibility of the derived orientation parameters.

\subsection{SGA galaxy measurements}
\label{subsec:allsga}
SGA provides a table of summary statistics for each imaged galaxy.\footnote{\url{https://portal.nersc.gov/project/cosmo/data/sga/2020/SGA-2020.fits}} From this sample of \sgatotal\ galaxies, we select all plausible spirals using the {\tt MORPHTYPE} column and selecting all columns with `S'. This gives a sample of \sgaselected\ disc-like galaxies. For each of these galaxies, we attempt to obtain the $grz$ imaging from the National Energy Research Scientific Computing Center (NERSC) portal.\footnote{\url{https://portal.nersc.gov/project/cosmo/data/}} 
We retain only galaxies for which the named galaxy is the central one in the image; in practice this means comparing the {\tt GALAXY} and {\tt GROUP\_GALAXY} column and discarding those entries for which there is no agreement. This leaves us with a sample of \sgaused\ galaxies. 

We provide a value-added catalogue\footnote{\url{https://github.com/ObservationalExpansions/Galaxy-Morphology/blob/main/SGA_catalogue/FLEX_measurements.h5}} for the \sgaused\ galaxies which includes 
\begin{enumerate}
    \item {\tt GALAXY}: the galaxy identifier.
    \item {\tt BA\_LEDA\_sga}, {\tt PA\_LEDA\_sga}, {\tt MORPHTYPE\_sga}: \textit{HyperLEDA}-provided parameters for each galaxy, which are used as the primary comparison in this paper.
    \item {\tt BA\_sga}, {\tt PA\_sga}, {\tt D26\_sga}: SGA-provided parameters for each galaxy, which can be compared in the same pipeline as \textit{LEDA}.
    \item {\tt ASCALE\_r}, {\tt ASCALE\_g} and {\tt ASCALE\_z}: the exponential scale length in each band, measured in pixels.
    \item {\tt eta\_g}, {\tt eta\_r}, {\tt eta\_z}: the $\eta$ value in each band (dimensionless).
    \item {\tt pa\_g}, {\tt pa\_r}, {\tt pa\_z}: the position angle in each band (in degrees), computed from $\vartheta$ (Equation~(\ref{PAeqn})).
    \item {\tt quality\_flag}: binary bit flag for indicating reconstruction quality.
\end{enumerate}
This catalogue can be cross-matched with the original SGA catalogue to compare all galaxy properties. 

\begin{figure*}
    \centering
    \includegraphics[width=\linewidth,height=0.9\textheight,keepaspectratio]{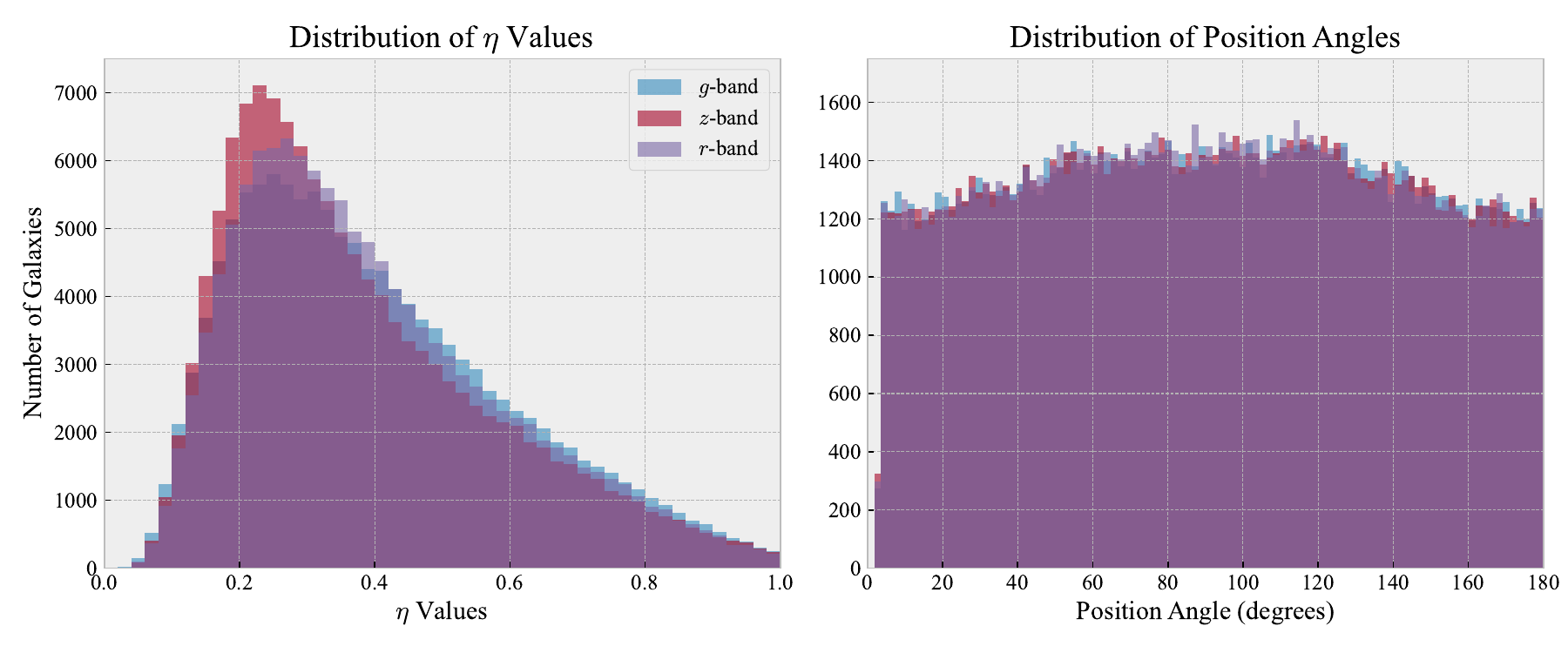}
    \caption{{\tt flex}-derived metrics for the SGA disc dataset. (i) Distribution of $\eta$ values between 0 and 1 for each band (left). (ii) Distribution of PA for each band (right).}
    \label{fig:etas_bands}
\end{figure*}

As an indication of the consistency between bands, we compare the $\eta$ and converted PA values in various bands. Figure~{\ref{fig:etas_bands}}~(i) and (ii) serves as a visual aid to interpret the consistencies across bands. Evident from the agreeing values in both measurements, the pipeline can be applied to any band. There is a clear distribution pattern for both $\eta$ and PA where the right-skewed $\eta$ distribution peaks around 0.25 and PA appears relatively Gaussian with a uniform peak near 90$^\circ$. The preferred band for our main results is the \textit{g}-band. 

\subsection{Estimated uncertainties}

By geometry, there is an intrinsic inclination and position angle for a disc galaxy. The dispersion of $\eta$ and PA values across \textit{g}, \textit{r}, and \textit{z} bands provides a direct estimate of measurement uncertainty. Since the orientation of the galaxy is invariant under the wavelength at which it is imaged, one should expect a minimal difference in the measurement across the band. 
For galaxies with \texttt{quality\_flag} less than or equal to three, we measure $\langle\eta_{\rm band~difference}\rangle=0.049$ and  $\sigma_{\eta_{\rm band~difference}}=0.085$, resulting in $\eta$ on average falling between $\pm0.025$ of each other band. When propagated to inclination, the uncertainty as a median is $\pm0.59^\circ$ degrees with a maximum of $\pm7.5^\circ$. This implies that the average galaxy in our dataset has strong agreement for the inclination measurement between bands, as anticipated.
Less than 1.5\% of the reduced sample show inter-band differences in $\eta$ large enough to produce a discrepancy of 0.3. The small subset of these cases that would incorrectly classify a galaxy as edge-on instead of face-on (or vice versa) are attributable to problematic images. Upon visual inspection, it appears that the highly discrepant sources tend to suffer from imaging artefacts in at least one of the bands. The \textit{z}-band is the most prone to such issues (e.g., glare streaks across the frame). Considering only the \textit{r} and \textit{g}-bands, only about 0.04\% of galaxies exhibit an inter-band difference in $\eta$ greater than 0.3, resulting in an average difference in inclination of $18.6^\circ$.
Tightening the flag constraint to \texttt{quality\_flag} $\leq 2$ or even $\leq 1$ further reduces the number of problematic galaxies to 34. However, in general, we retain the \texttt{quality\_flag} $\leq 3$ criterion to preserve the statistical power of the larger sample.

Applying the same logic to PA across bands, we find a median difference between any arbitrary two bands to be $\langle \Delta\text{PA}\rangle=2.15^\circ$, with a standard deviation of $\sigma_{\text{PA}}=8.89$. Values indicating PA values close to $180^\circ$ are theoretically equivalent to a PA of $0^\circ$. To minimise outliers in bands close to the extreme angles, a smaller range between $15^\circ$ and 165$^\circ$ was analysed, resulting in a statistical sample of 99,680 galaxies. After accounting for the imaging problems to which the \textit{z}-band is most susceptible, the fraction of galaxies with PA differences greater than $15^\circ$ decreases from 7.88\% to 2.55\%.
Differences from reported literature values are thus unlikely to stem from large internal uncertainties within our pipeline, and are instead likely a reflection of the differences in methodology and assumed priors. 

%% file: 5Results.tex
The \textit{HyperLEDA} database is a comprehensive compilation of galaxy parameters including morphological classifications, photometry, kinematics and structural properties derived from a range of observational sources. Inclination and PA values in \textit{HyperLEDA} are typically estimated from the observed axial ratio (\textit{b/a}) using empirical assumptions about the intrinsic thickness of galactic discs. While this approach has served as a standard technique, it is subject to systematic uncertainties, particularly for galaxies with non-axisymmetric structures, bars, or irregular morphologies. We employ the \textit{HyperLEDA} catalogue and the reevaluated \citet{GalaxyInclPaper} as one of the references for measurements to evaluate our values derived from the Fourier-Laguerre expansions. \citet{GalaxyInclPaper} forms the second gauge for inclinations for the `golden sample'.
By analysing the comparisons, we assess the accuracy, reproducibility, and scalability of the new approach. The following sections outline the validation of our technique by a worked example, later using the carefully selected `golden sample' of galaxies, followed by the application to the full SGA dataset.

\subsection{Worked example: NGC3067}
\label{subsec:ngs3067}
\begin{figure}
    \centering
    \includegraphics[width=1\linewidth]{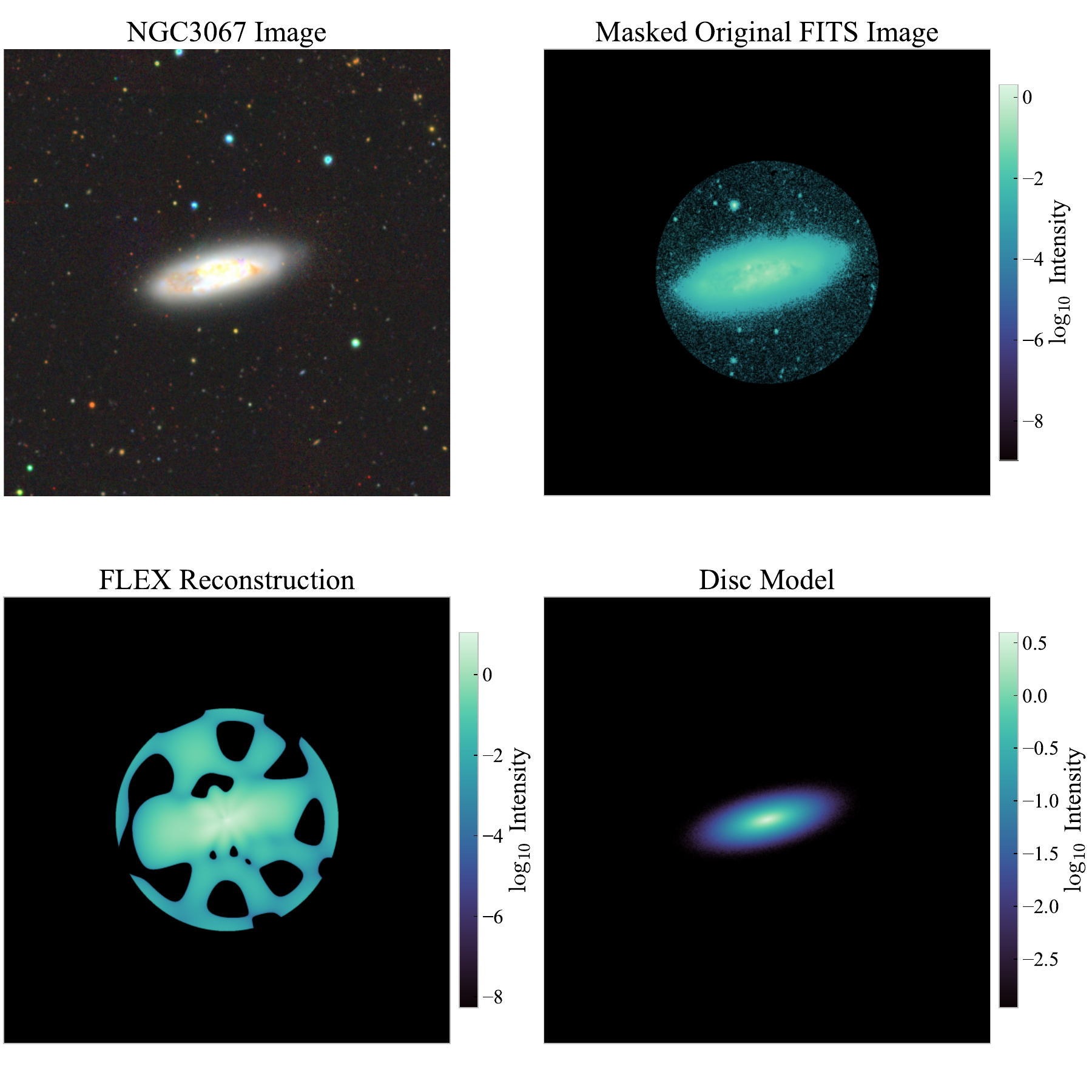}
    \caption{(i) $grz$ image of NGC3067 via SGA (upper left), (ii) $\log_{10}$ of intensity of $g$-band masked image of galaxy NGC3067 (upper right), (iii)
    $\log_{10}$ of intensity of Fourier-Laguerre expansion of the image under the same colour map and mask (lower left) and (iv) $\log_{10}$ of intensity of galaxy with inclination of 68.6$^\circ$ and a PA of $104^\circ$ generated by \texttt{discmodel} (lower right).}
    \label{fig:panel}
\end{figure}

Following the pipeline with one galaxy from the catalogue, we will find the relevant angles and verify them beginning with a Fourier-Laguerre expansion of the image. The galaxy NGC3067 from the SDSS catalogue pictured in Figure~\ref{fig:panel}~(i)\footnotemark\footnotetext{NGC3067 image ($grz$ composite) downloaded from NERSC SGA portal: 
\url{https://portal.nersc.gov/project/cosmo/data/sga/2020/data/149/NGC3067/NGC3067-largegalaxy-image-grz.jpg}. 
Data originally from SDSS DR14.} is a barred disc galaxy, which itself has been studied with respect to its halo gas \citep{1989Natur.338..134C,Tumlinson_1999}. The image is expanded with  Fourier-Laguerre pairs with the \texttt{galaxymorphology function}. Figure~\ref{fig:panel}~(ii) and (iii) demonstrate the expansion and its detail capturing the shape of the disc and the centre structure of the masked galaxy in terms of the intensity of the pixels. 
The Fourier-Laguerre expansion returns a matrix of the form described in Equation~(\ref{matA}). Sine and cosine components of the matrix elements (the sine and cosine \textit{coefficients}) are substituted into Equation~(\ref{eta}) to find the metric $\eta$. For this particular galaxy, a value of $\eta_g=0.61$  is returned. Implementing the numerics and inversion as a result of Equation~(\ref{eta_function}) to find the corresponding inclination with the recommended fit coefficients, an inclination of $i=68.6^\circ$ is produced. According to the \textit{HyperLEDA} catalogue, galaxy NGC3067 has an inclination of $81.6^\circ$ and to \citet{GalaxyInclPaper}, a validated inclination of $71^\circ$. Our value of 68.6$^\circ$ lies closer to the updated value in the recent literature. While only one galaxy, we have not cherry-picked a particularly good fit and will demonstrate below that {\tt flex}-derived measurements are an improvement on axial ratio measurements.

Using Equation~(\ref{PAeqn}) and substituting the corresponding matrix elements decomposed as sine and cosine from Equation~(\ref{vartheta}), a PA of $103.96^\circ$ in the \textit{g}-band is produced. As stated in the \textit{HyperLEDA} catalogue, this galaxy has a PA of $104.1^\circ$, a value which differs from our result by less than one degree. 

In order to visualise the accuracy of these measurements, we compare the original galaxy with simulated a disc galaxy made with \texttt{discmodel} using the orientation angles found above. Figure~\ref{fig:panel}~(iv) draws similarities between the studied galaxy and the disc with its measured orientation. This disc model is produced with the scale length derived from the noise analysis of the original image and the mass derived as a sum of the remaining pixels after the mask.
Despite differences in the distribution of the intensity of the pixels and a lack of spirals or bars in the disc model, it is clear that the elliptical shape, inclination and rotation, produced by the  measurements taken with the distinct metrics maintains the orientation of the original galaxy. 

\subsection{Golden sample results}\label{subsec:gsres}

\begin{figure}
    \centering
    \includegraphics[width=1\linewidth]{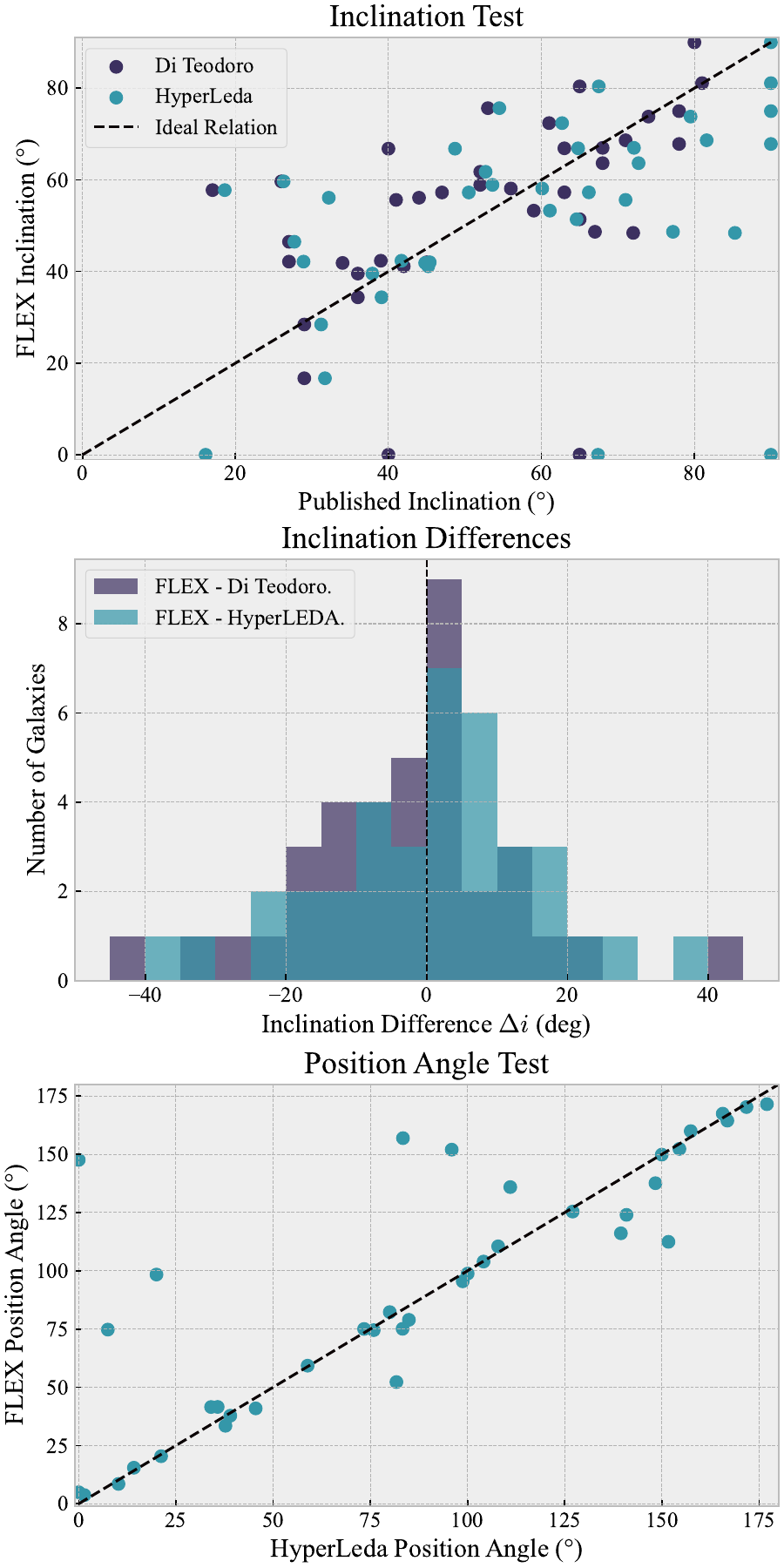}
    \caption{(i) {\tt flex}-derived inclination value from $\eta$ against published inclinations (top). Each galaxy appears with two $x$-axis values: once with the inclination drawn from the \textit{HyperLEDA} database (light markers) and once with the inclination drawn from \citet{GalaxyInclPaper} (dark markers). (ii) Distribution of differences in literature inclination with {\tt flex}-derived inclination (middle). (iii) {\tt flex}-derived PA against \textit{HyperLEDA} PA values (bottom).}
    \label{fig:3_paperandleda}
\end{figure}

\begin{figure}
    \centering
    \includegraphics[width=0.9\linewidth]{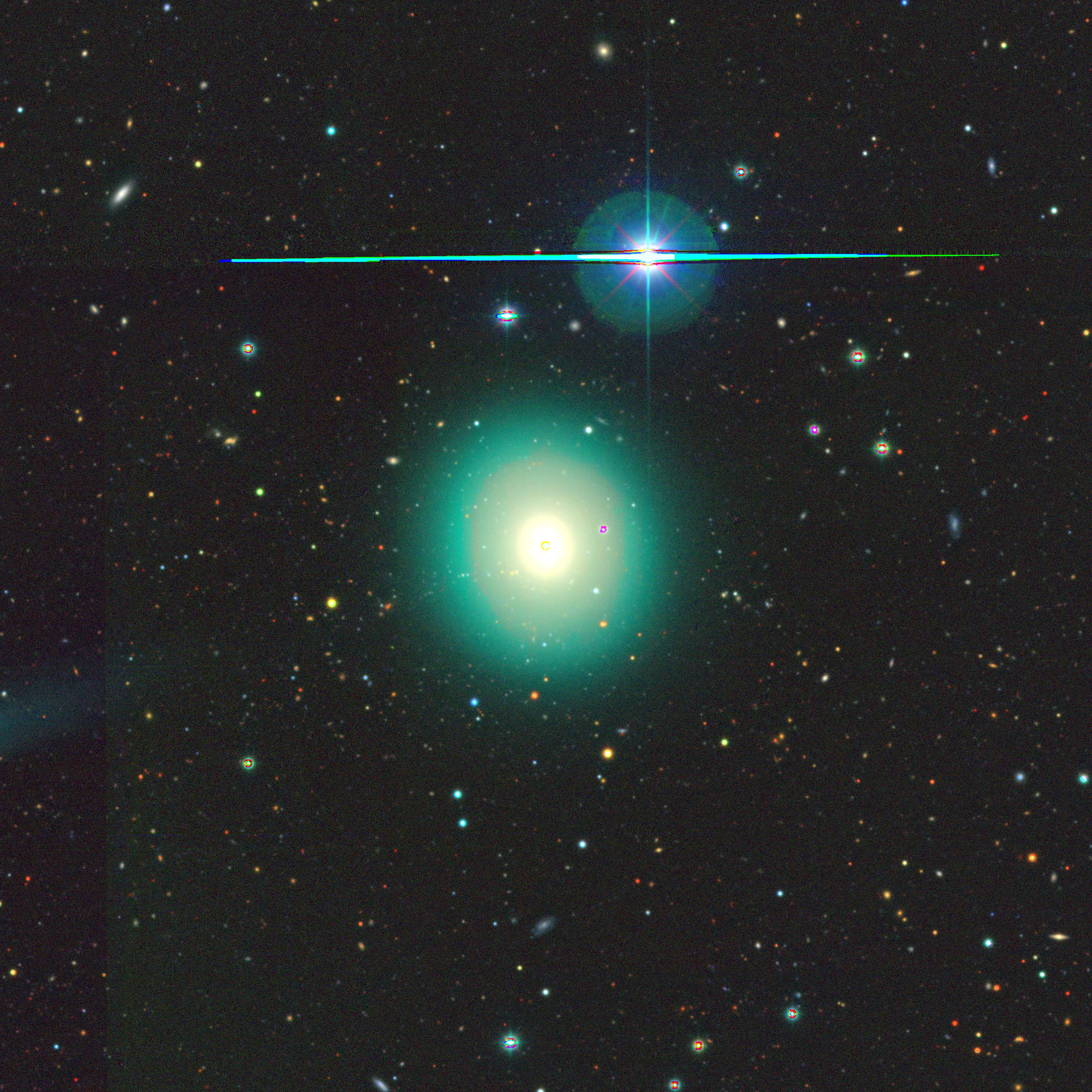}
    \caption{NGC4203 image ($grz$ composite) showing a nearly face-on galaxy ($i_{\rm flex}\approx0$), while literature values report $i_{\rm literature}=65-90^\circ$. Image from NERSC SGA portal: \url{https://portal.nersc.gov/project/cosmo/data/sga/2020/data/183/NGC4203/NGC4203-largegalaxy-image-grz.jpg}.}

    \label{fig:4203}
\end{figure}

Implementing the recommended inversion relation to recover inclination and PA in the \textit{g}-band, we compare them with the literature. Figure~\ref{fig:3_paperandleda}~(i) and (iii) display the angle values, while Figure~\ref{fig:3_paperandleda} (ii) presents the distribution of the difference in values between our \texttt{flex}-derived inclination and \textit{HyperLEDA}. The PA values agree for this small sample and the inclination values sit close to the ideal 1:1 line. The median difference in values between our inclinations and \citet{GalaxyInclPaper} is 10.0$^\circ$ and between our values and \textit{HyperLEDA} is 9.7$^\circ$. Arguably, the difference is smaller when comparing our values with the values presented by \textit{HyperLEDA}, however, there are a number of inclination measurement catastrophic failures in this database: for example, NGC4203.
\textit{HyperLEDA} inclination values call NGC4203 an edge-on galaxy ($i=90^\circ$). \citet{GalaxyInclPaper} publishes an inclination of $65^\circ$. Our algorithm implemented in all three bands returns an inclination of $\approx0^\circ$. Given that this is an image with substantial potential contamination around the galaxy which could be interpreted as a fully edge-on galaxy (see Figure \ref{fig:4203}), it is possible that this has affected earlier measurements. 
Once the two outliers affecting extreme disagreeing values between \citet{GalaxyInclPaper} and \textit{HyperLEDA}, or visibly flawed characterisations of a galaxy have been considered, our inclinations lie closer to the \citet{GalaxyInclPaper} inclinations, differing only by 1 degree on average. Assessing the differences in PA measurements with \textit{HyperLEDA} returns a median PA difference of 3.27$^\circ$.

\subsection{All SGA galaxies}

\begin{figure}
    \centering
    \includegraphics[width=1\linewidth]{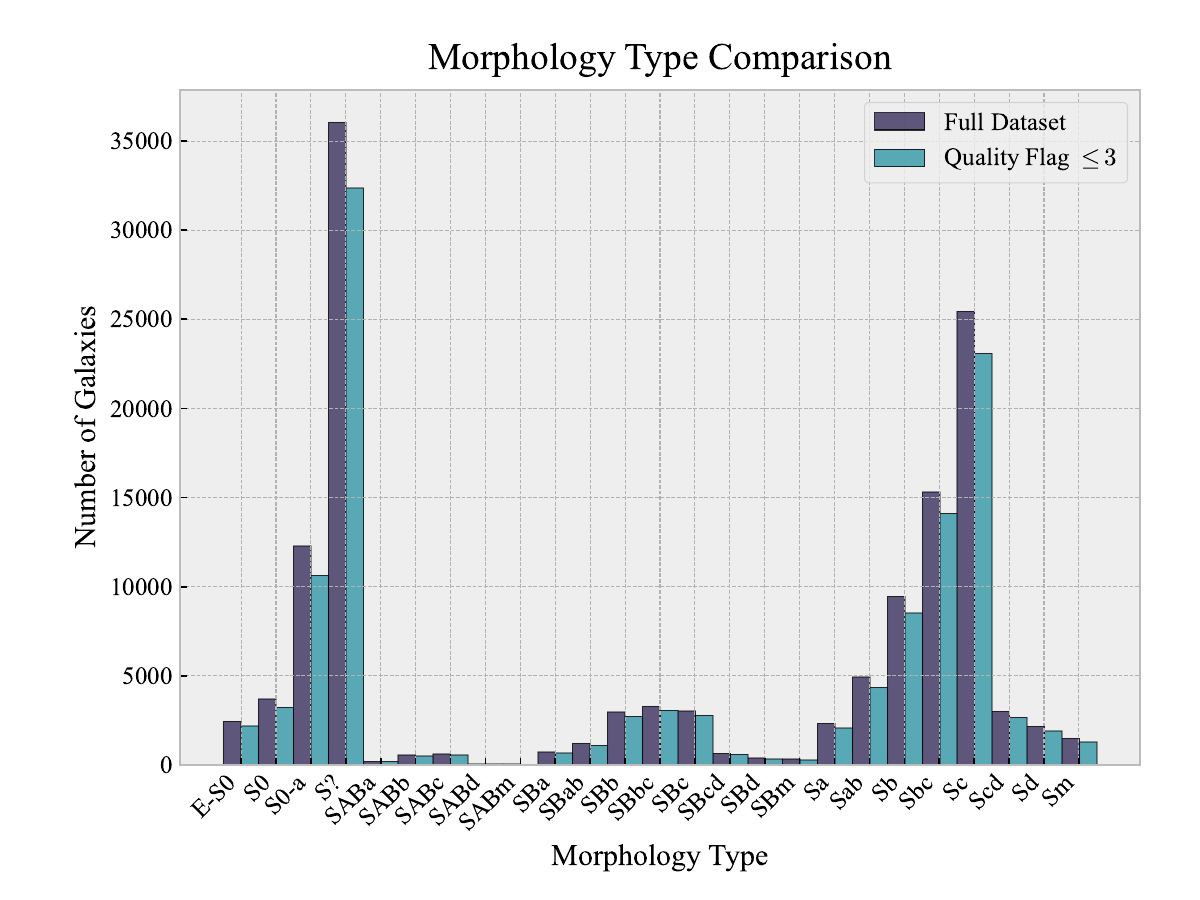}
    \caption{Initial morphology distribution (dark bars) and quality flag reduced set morphology distribution (light bars) in the SGA sample of disc galaxies.}
    \label{fig:morph_distribution} 
\end{figure}

The full SGA disc dataset consists of \sgaselected\ galaxies. We show the morphology types of each galaxy in Figure~\ref{fig:morph_distribution}. Using the derived quality metric, we identify \sgaflag\ galaxies from the initial set of \sgaselected\ galaxies as ones with quality flags of three or less, as outlined in Section~\ref{subsec:allsga}. The morphology distribution of the remaining galaxies is shown alongside the initial input sample in Figure~\ref{fig:morph_distribution}. There is a wide variety of spiral galaxies, and the whole set remains represented and symmetric without systematic biases in the quality flag-selected set.

\begin{figure}
  \centering
  \includegraphics[height=0.89\textheight,keepaspectratio]{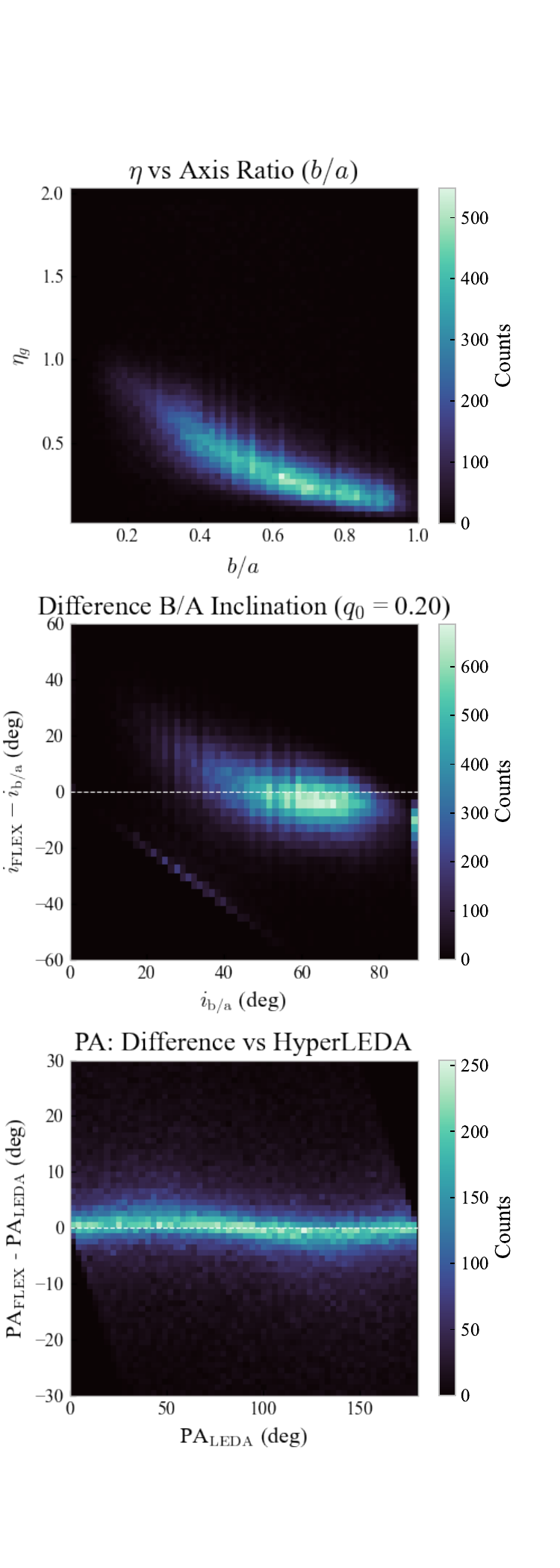}
  \caption{(i) Comparison of the $\eta$ metric derived from the quality reduced sample of galaxies with the given \textit{b/a} ratio (top). (ii) Difference between processed inclinations from the $\eta$ values via recommended inversion and approximate inclination values derived directly from the \textit{b/a} ratio against the \textit{b/a} ratio derived inclinations themselves (middle). (iii) Difference in PA measured from coefficients and {\tt PA\_LEDA} column in dataset against {\tt PA\_LEDA} column in dataset (bottom).}
    \label{fig:ba_inc_SGA}
\end{figure}

Figure~\ref{fig:ba_inc_SGA}~(i) presents a strong correlation between $\eta$ and \textit{b/a}, which both convert to an inclination. The \textit{b/a} values were converted into an inclination with the standard formula derived with Hubble's findings \citep{Hubble}, picking an average $q_0$ value of 0.2 (the estimated intrinsic axial ratio of the galaxy viewed edge-on). Applying the recommended conversion parameters to $\eta$ described in Appendix~\ref{Appendix_recommendedIncl} results in the comparison in Figure~\ref{fig:ba_inc_SGA}~(ii). A clear correlation is visible emphasising the abilities of this pipeline. A small percentage of the sample appears detached from the majority of the data points. With a tendency to over-estimate lower inclinations with this particular inversion relation, there are some galaxies which our pipeline considers to be directly face on while the \textit{b/a} values give a low inclination. Testing with different bands than \textit{g} also maintains a similar trend. 

The full distribution of inclinations recovered for the full set of SGA galaxies which were analysed peaks near 60$^\circ$, consistent with expectations for randomly oriented discs \citep{1993ApJ...402...15L,article60deg,60deg10.1093/mnras/stv1948}.

Taking into consideration the variety of morphologies in this set from Figure~\ref{fig:morph_distribution}, we can conclude that this method is reliable across classifications. Computing the average difference between the {\tt flex}-derived inclination and the value derived from the \textit{b/a} ratio with $q_0=0.2$, a median difference is found to be $6.9^\circ$.

Comparing the PA values from \textit{HyperLEDA} to our measured PA in the \textit{g}-band, Figure~\ref{fig:ba_inc_SGA}~(iii), displays the strong correlation between the measurements. Computing an average difference directly, omitting those greater than 165$^\circ$ and below 15$^\circ$ due to the periodicity 180$^\circ$ beyond 0$^\circ$,  a median difference of $4.09^\circ$ is returned.

\subsection{Orientation-determination caveats}

The uncertainties in the {\tt flex}-derived inclination and PA values arise primarily from three sources: (i) the finite truncation of the Fourier–Laguerre series, (ii) noise and contamination within the imaging data, and (iii) the adopted calibration between the dimensionless coefficient ratio $\eta$ and inclination. The truncation of the series at $n_{\mathrm{max}}=10$ was chosen as a compromise between accuracy and computational efficiency. Image noise and contamination effects were assessed through the quality‐flag procedure defined in Section~\ref{subsec:allsga}. While some image contamination is unavoidable, comparison across bands for the small percentage of disagreeing $\eta$ values allows for large errors to be mitigated. Accepting an average of the inclinations from the small sample of disagreeing measurements serves as a balance until a robust selection system is developed in the future. 

When determining inclination with this technique, the largest source of systematic uncertainty remains the empirical calibration of the $\eta$–inclination relation; recalibration using alternative datasets or bandpasses could shift inclinations. Considering the total set of SGA inclinations derived with the traditional approach applies an approximated value for $q_0=0.2$, we will use the `golden sample' \textit{HyperLEDA} and \citet{GalaxyInclPaper} measurements as an uncertainty gauge in their difference from our measurements. These comparisons indicate typical inclination uncertainties of roughly 10$^\circ$. The close agreement between the uncertainty distributions in the golden sample and in the full dataset suggests that the golden sample is statistically representative of the larger set in both PA and inclination. To a significant figure, the uncertainty in the PA is 5$^\circ$ in both the full set and subset.

Physically, an aspect which can contribute to poor measurements across methods is the assumption that the disc itself is razor thin. Studies and observations show that mass distribution does  not lie in the same plane, however it is often a reasonable assumption at the scale of a galaxy \citep{mass_2004A&A...418L..27B}. There may be cases for which this lopsidedness cannot be neglected which impacts the measurements, but this effect would impact any approach which takes advantage of the two-dimensional ellipticity of disc galaxies. Similarly, when a particular galaxy is viewed face-on, the `perfect circle' assumption may not always be robust \citep{refId0}. This means that fitting an ellipse to a skewed shape may mischaracterise the orientation of the galaxy. A skewed shape would also be present in the expansions used here, and future work may involve a \texttt{discmodel} with non-exponential mass distribution or parameters corresponding to fine-tuning the delicate structure of the galaxies and their substructure such as bars and spirals.